\documentclass[usenatbib]{mn2e}
\usepackage{graphicx}

\usepackage{amssymb}
\usepackage{color}
\usepackage{float}

\newcommand{\bn}{\begin{enumerate}}
\newcommand{\en}{\end{enumerate}}
\newcommand{\bi}{\begin{itemize}}
\newcommand{\ei}{\end{itemize}}

\def\gtorder{\mathrel{\raise.3ex\hbox{$>$}\mkern-14mu
    \lower0.6ex\hbox{$\sim$}}}
\def\ltorder{\mathrel{\raise.3ex\hbox{$<$}\mkern-14mu
    \lower0.6ex\hbox{$\sim$}}}

\newcommand{\apj}{ApJ}

\newcommand{\apjl}{ApJL}

\newcommand{\mnras}{MNRAS}

\title[Dark Matter Bars in Spinning Halos]{Dark Matter Bars in Spinning Halos} 
\author[Angela Collier, Isaac Shlosman, and Clayton Heller]
{Angela Collier$^{1}$\thanks{E-mail: angela.collier@uky.edu},
Isaac Shlosman$^{1,2}$\thanks{E-mail: shlosman@pa.uky.edu},
Clayton Heller$^{3}$
\\
\footnotemark    
$^{1}$ Department of Physics \& Astronomy, University of Kentucky, Lexington, KY 40506-0055, USA\\
$^{2}$ Theoretical Astrophysics, Graduate School of Science, Osaka University, Osaka 560-0043, Japan\\  
$^{3}$ Department of Physics \& Astronomy, Georgia Southern University, Statesboro, GA 30460, USA \\
}

\begin{document}

\date{Accepted ?; Received ??; in original form ???}


\maketitle

\begin{abstract}
We study nonlinear response of spinning dark matter (DM) halos to dynamic and secular evolution of stellar bars in the embedded galactic disks, using high-resolution numerical simulations. For a sequence of halos with the cosmological spin parameter $\lambda=0-0.09$, and a representative angular momentum distribution, we analyze evolution of induced DM bars amplitude and quantify parameters of the response as well as trapping of DM orbits and angular momentum transfer by the main and secondary resonances. We find that (1) maximal amplitude of DM bars depends strongly on $\lambda$, while that of the stellar bars is indifferent to $\lambda$; (2) Efficiency of resonance trapping of DM orbits by the bar increases with $\lambda$, and so is the mass and the volume of DM bars; (3) Contribution of resonance transfer of angular momentum to the DM halo increases with $\lambda$, and for larger spin, the DM halo `talks' to itself, by moving the angular momentum to larger radii --- this process is maintained by resonances; (4) Prograde and retrograde DM orbits play different roles in angular momentum transfer. The `active' part of the halo extends well beyond the bar region, up to few times the bar length in equatorial plane and away from this plane. (5) We model evolution of diskless DM halos and halos with frozen disks, and found them to be perfectly stable to any Fourier modes. Finally, further studies  adopting  a range of mass and specific angular momentum  distributions of the DM halo will generalize the dependence of DM response on the halo spin and important implications for direct detection of DM and that of the associated stellar tracers, such as streamers.
\end{abstract}


\begin{keywords}
methods: dark matter --- methods: numerical --- galaxies: evolution, galaxies: formation --- galaxies: interactions --- galaxies: kinematic \& dynamics
\end{keywords}

\section{Introduction}
\label{sec:intro}

Self-gravitating systems still challenge our understanding. Stellar bars in disk galaxies can form either spontaneously, as a result of the bar instability \citep[e.g.,][]{hohl71,sell80,atha92,bere98}, or following interactions with other galaxies \citep[e.g.,][]{toom72,geri90,bere04}, or in stellar disk interactions with halo's dark matter (DM) clumps \citep{rom08}. Evolution of stellar bars has been studied and analyzed in numerical simulations in nonrotating halos \citep[e.g.,][]{atha83,sell87,bere98,dub09}. Theoretical works predicted a minor effect of spinning parent halos on the embedded stellar disks evolution \citep{wein85}.  Contrary to these expectations, recent results indicate that a halo spin has a profound effect on the stellar bar evolution, and affects angular momentum redistribution in disk-halo systems \citep{saha13,long14,coll18}.  

In previous works, we have focused on the evolution of stellar bars in spinning halos and the associated angular momentum transfer in the disk-halo system \citep{long14,coll18}. Here we aim to analyze the parent DM halo response to this process. While it is known that DM response is triggered by the formation of a bar in the disk \citep{tre84,wein85,atha05,shlo08,atha13,pete16}, its properties in spinning halos are largely unknown.

Models of disk galaxies in spinning halos brought up a number of surprises. First, the bar instability appears to
accelerate compared to the nonrotating halos \citep{saha13,long14}. Second, the vertical buckling instability in the
bar is more profound, and weakens the bars progressively with increased halo spin. Defining the halo spin as
$\lambda=J/J_{\rm K}$, where $J$ and $J_{\rm K}$ are the halo angular momentum and its maximal angular momentum,
respectively \citep[e.g.,][]{bull01}, the important transition range lies in $\lambda\sim 0.03-0.06$ \citep{coll18}.
Above $\lambda\gtorder 0.03$, the stellar bar amplitude has difficulty to recover after buckling even over the secular
time of evolution. Consequently, the bar braking ability against the DM becomes dramatically weaker. Moreover, for
$\lambda\gtorder 0.06$, the bars essentially dissolve, leaving behind a weak oval distortion. Thus, bar evolution in
spinning halos can affect a substantial fraction of disk-halo systems.

Works that did not reach similar conclusions on the importance of halo spin to bar dynamics have either limited their analysis to $\lambda\ltorder 0.03$ halos \citep{pete16}, and hence did not test the relevant range in $\lambda$, or limited the evolution time to the pre-buckling stage of the bar instability \citep{saha13}. Moreover, cosmological simulations that include a broader range of $\lambda$ and secular evolution of disks do not have sufficient resolution as achieved in studies of isolated halos. Their treatment of angular redistribution between the DM halos and embedded disks are not precise enough and hence can miss this effect altogether. 

To fully understand the ramification of this new effect, we need to analyze its dependence on two important
dynamical indicators --- the distributions of mass and angular momentum in the halo. The former has a universal
character \citep[e.g.,][]{nava96} with various degrees of central mass concentration, which reflects the formation
time. Baryons tend to increase the central concentration.

Distribution of angular momentum in the halo has been claimed to be universal as well for pure DM halos
\citep[e.g.,][]{bull01}. The addition of baryons can  modify this distribution in principle by increasing the angular
momentum within the inner halo, which can be measured from halos, for example, in Illustris simulation
\citep{vogel14}. We note that the model halos of \citet{coll18} used in the present work are representative of $J$ distribution in baryonic halos
and lie within one $\sigma$ from its median. The detailed study of bar dynamics for a range in $J$ and mass
distributions in DM halos is forthcoming.

The basic questions that must be answered when dealing with the DM bar evolution in spinning halos are as follows. Is the strength of a DM bar affected by the halo spin $\lambda$? Are DM bar mass and shape dependent on the halo spin? What type of orbits comprise the DM bar? Are all these orbits absorbing the angular momentum from the disk? What part of the halo absorbs the angular momentum from the disk, and how does the efficiency of DM orbit trapping by the resonances depend on the halo spin? We aim to resolve these questions.

In the previous work, we have confirmed that the timescale of the stellar bar instability is shortened along the $\lambda$ sequence, as shown by \citet{saha13} and \citet{long14}. While the maximal strength of these bars is independent of $\lambda$, their secular
evolution depends strongly on the parent halos spin. Probably the most interesting and unexpected effect of stellar
bar evolution in spinning halos is that the buckling instability of the bar is more destructive with $\lambda$, and
stellar bars have have increasing difficulty to recover their strength after buckling. This transition occurs in the
range of $\lambda\sim 0.03-0.06$. Close to the upper value and above it, the stellar bars are basically destroyed by
the buckling instability and never regrow. Models with spherical, oblate and prolate halos have been run and behaved
similarly. The DM response to the underlying stellar bar perturbation dies out immediately with its disappearance was
shown for the first time in \citet{shlo08}. 

Observational corollaries of this evolution include decreasing braking of a stellar bar against the DM, and much higher ratios of corotation-to-bar size, $r_{\rm CR}/r_{\rm bar} > 2$ after buckling, well beyond the ratios encountered in $\lambda=0$ halos, $r_{\rm CR}/r_{\rm bar}\sim 1.2\pm 0.2$ \citep{atha92}. Furthermore, stellar bar growth experiences difficulties with increasing $\lambda$, and saturates completely for $\lambda\gtorder 0.05$.  The high-$\lambda$ halos anti-correlate with the existence of {\it ansae}, and exhibit smaller size and mass of the peanut/boxy-shaped bulges. 

That angular momentum $J$ flows from a barred stellar disk to a DM halo is known for quite some time \citep[e.g.,][]{sell80,wein85,debat00,atha03,atha05,marti06,bere07,wein07a,wein07b,dub09}. That this flow is mediated by the orbital resonances, and especially by the inner Lindblad resonance (ILR), outer Lindblad resonance (OLR), and the corotation resonance (CR) has been established as well \citep{lynd72,wein85,atha03,marti06}.

Importantly, the parent halo spin strongly affects the angular momentum transfer between the disk and its halo, but this point is required to be investigated further \citep{coll18}. One expects that the DM halo orbital structure will help to understand the intricacies of angular momentum flow in the system, many of which remain unclear. \citet{wein85} has suggested that the increase in the number of prograde particles in the spinning halo (with respect to disk rotation) increases naturally the fraction of halo particles trapped by the main resonances and speeds up the bar instability. But this assumption was never verified in a quantitative analysis. Nor was it verified how the trapping by resonances explains the secular evolution of stellar bars in spinning halos. We attempt to tackle these issues in this paper and in \citet{coll19b}.

To quantify the angular momentum flow in the disk-halo systems, one can take a dual approach. It is possible to follow the rate of angular momentum flow with the method designed by \citet{villa09} \citep[see also section\,\ref{sec:spectral};][]{villa10,long14,coll18}. To determine the role of the resonances in this transfer, we refer to the orbital spectral analysis \citep[e.g.,][]{binn82,atha03,marti06,dub09}. This method allows us to determine the fraction of DM orbits trapped by the resonances. One can apply this method in frozen potentials and integrate the orbit for a fixed and large number of periods. Alternatively, one can do this in the live potential of the system. This, however, has its disadvantages --- the number of time periods to integrate along the orbit will be small, leading to unreasonable widening of the resonances. Hence, we follow the former method and use it in order to find the DM orbit trapping efficiency by the resonances as well as amount angular momentum transported by these resonances, comparing it along the $\lambda$ sequence. 

Because we focus on properties of DM halos with increasing spin, one should ask whether diskless halos with the same $\lambda$ are stable against spontaneous breaking of the axial symmetry.

Stability of pure diskless halos with a non-zero cosmological spin is subject to diverging opinions. Based on \citet{jeans19} theorem, \citet{lynd60} has argued that spherical halos with all particle tangential velocities reversed in the same direction are stable. Of course, Jeans theorem does not capture the elusive bar instability, when the system can lower its energy by breaking the axial symmetry, as in Maclaurin sequence of rotating ellipsoids \citep[e.g.,][]{chand69,binn08}. Hence, \citet{alle92} claimed, based on their numerical simulations, that rotating models of spherical, oblate, and prolate $N$-body systems become unstable and form "triaxial bars." However, \citet{sell97}, after reproducing their initial conditions, found that this instability resulted from an error in the code, and newly re-run models were completely stable. Though these authors warned that a fast spinning halo --- one with all orbits rotating in the same direction, may still become bar unstable. 

Furthermore, \citet{dub94} cautioned against using halos of large spin after seeing bar formation in oblate Evan's model systems of $\lambda = 0.18$. However, the range of $\lambda$ used in our work is much lower,
$\lambda\ltorder 0.09$, and we do not expect that our halos are unstable. Nevertheless, we test diskless spherical halos with maximal spin which can be obtained in our models, up to $\lambda = 0.108$, in section\,\ref{sec:pureDM}. Moreover, we tested these halos with an embedded {\it frozen} disk, to account for changes that can be introduced by the disk gravitational potential. All our diskless halos are stable against bar instability or any other global instability over the time of 10\,Gyr. Hence, we find that limiting the spin to $\lambda=0.03$, as motivated by \citet{pete16} is not warranted. We have limited our analysis to only spherical models. Oblate and prolate halos modeled by \citet{coll18} will be discussed elsewhere. 

This paper is structured as follows. Section\,2 deals with numerics, including initial conditions and orbital spectral analysis. Section\,3 presents our results, starting with diskless DM halos and switching to disk-halo systems. Section\,4 discusses our results and theory corollaries, and we end with conclusions.

\section{Numerics}
\label{sec:numeric}

\subsection{Model Setup and Initial Conditions}
\label{sec:ICs}

We analyze models of disk-halo systems described in \citet{coll18}, and additional models of isolated DM halos. Our pure DM halo models and disk-halo models differ only with the DM halo spin, $\lambda$. The initial conditions have been created using a novel iterative method \citep[][]{rodio06,rodio09,long14,coll18}. These models have been run using the $N$-body part of the GIZMO code originally described in \citet{hop15}. We choose units of distance and mass as 1\,kpc and $10^{10}\,M_\odot$, respectively. This leads to the time unit of 1\,Gyr. The DM halos have been modeled with $N_{\rm h} = 7.2\times 10^6$, and stellar disks with $N_{\rm d} = 8\times 10^5$. So the ratio of masses of DM particles to stellar particles is close to unity. For convergence test, we run models with twice the number of particles and obtained similar evolution.

Halo shapes include spherical, oblate and prolate models, but only the former ones are discussed here. The halo spin has been varied by inverting a fixed fraction of tangential velocities for retrograde DM particles (with respect to the disk rotation), which does not change the solution of the Boltzmann equation \citep{lynd60,wein85,long14,coll18}. $J$ of each halo has a log-normal universal distribution \citep{bull01}.

The models contain an exponential disk with density:

\begin{eqnarray}
\rho_{\rm d}(R,z) = \bigl(\frac{M_{\rm d}}{4\pi h^2 z_0}\bigr)\,{\rm exp}(-R/h) 
     \,{\rm sech}^2\bigl(\frac{z}{z_0}\bigr).
\end{eqnarray}
Here $M_{\rm d}=6.3\times 10^{10}\,M_\odot$ is the disk mass, the radial scalelength is $h=2.85$\,kpc, and the scaleheight is $z_0=0.6$\,kpc. $R$ and $z$ are cylindrical coordinates. 

The halo density is given by \citet[][hereafter NFW]{nava96},

\begin{equation}
\rho_{\rm h}(r) = \frac{\rho_{\rm s}\,e^{-(r/r_{\rm t})^2}}{[(r+r_{\rm c})/r_{\rm s}](1+r/r_{\rm s})^2}
\end{equation}
where $\rho(r)$ is the DM density in spherical coordinates, $\rho_{\rm s}$
is the (fitting) density parameter, and $r_{\rm s}=9$\,kpc is the characteristic radius, where the power 
law slope is (approximately) equal
to $-2$, and $r_{\rm c}=1.4$\,kpc is a central density core. We used the Gaussian cutoffs at 
$r_{\rm t}=86$\,kpc for the halo and $R_{\rm t}=6h\sim 17$\,kpc
for the disk models, respectively. The halo mass is $M_{\rm h} = 6.3\times 10^{11}\,M_\odot$, its central mass
concentration $c\sim 90\,{\rm kpc}/r_{\rm s}\sim 10$, and 
halo-to-disk mass ratio within $R_{\rm t}$ is 2. 

We follow the notation of \citet{coll18} to abbreviate the disk-halo models, namely, $P$ for prograde spinning halos, followed by the value of $\lambda$ multiplied by 1,000. The Standard Model is defined as that of a non-rotating spherical halo, P00. Pure DM halo models are denoted as $H$, followed by $1,000\lambda$, as in disk-halo models.  More details can be found in Collier et al.

\subsection{Orbital Spectral Analysis}
\label{sec:spectral}

To examine the role of resonances in angular momentum transfer within the disk-halo systems, we use the orbital spectral analysis method \citep{binn82,atha03,marti06,dub09}. We perform the orbit spectral analysis in the frozen potential. The bar pattern speed is fixed in time as well. All spiral features that have the same pattern speed as the bar, are taken into account.

Our goal here is to quantify the role of the important resonances in angular momentum transfer by statistically sampling the stellar disk and DM halo orbits in our simulations and capture the angular velocity, $\Omega$, and radial epicyclic frequency, $\kappa$, for individual orbits, and gain insight into the resonance structure as a whole. Additional frequency, $\Omega_{\rm bar}$, is the bar pattern speed measured for each model at the time when the potential is frozen.  

We evolve stellar and DM test particles in frozen gravitational potential of the system at time $t$, for 50 orbits. By integrating the test particles trajectories, we record the corresponding cylindrical coordinates, $r$, $\phi$ and $z$. For each orbit, we determine the power spectrum of the representative frequencies, $\Omega$ and $\kappa$, by applying the Fourier transform to the $\phi$ and to $r$ values of each particle along every time step in their orbit. Next, we calculate the orbit fraction as a function of normalized dimensionless frequency $\nu\equiv (\Omega-\Omega_{\rm bar})/\kappa$, binned in $\Delta\nu=0.01$. We choose a sample of DM and stellar particles, 200,000 each, for this analysis. The test particles were randomly chosen from the entire sample of particles. The main resonances of interest are the inner Lindblad resonances, ILR, the corotation resonance, CR, and the outer Lindblad resonance, OLR. They correspond to $\nu=0.5$, 0, and -0.5, respectively.

The code is automated to capture thousands of orbits simultaneously. Test particles are identified and sorted by radius, in order to group them in bins of a similar dynamical time. This step is parallelized  and run using schwimmbad \citep{TheSchwimmbad} --- an MPI tool for Python.

\section{Results}
\label{sec:results}

We start with basic analysis of diskless DM halos in order to verify their stability, and continue with halos hosting stellar disks. To calibrate our simulations we run a number of new models of spherical diskless DM halos within the same range of $\lambda\sim 0-0.09$. Moreover, we have added three additional models: diskless halos with $\lambda=0.10$ and 0.1077, as well as P90 model with frozen stellar disk. The reasons for these additional models are explained in Section \ref{sec:pureDM}. We also measure the Fourier amplitude of developing DM bars in response to evolving stellar bars, present the results of the orbital spectral analysis, and determine the rates of angular momentum transfer in the disk-halo systems, emphasizing the role of prograde and retrograde DM orbits.

\subsection{Diskless Spinning DM Halo Systems}
\label{sec:pureDM}

\begin{figure}
\centerline{
 \includegraphics[width=0.55\textwidth,angle=0] {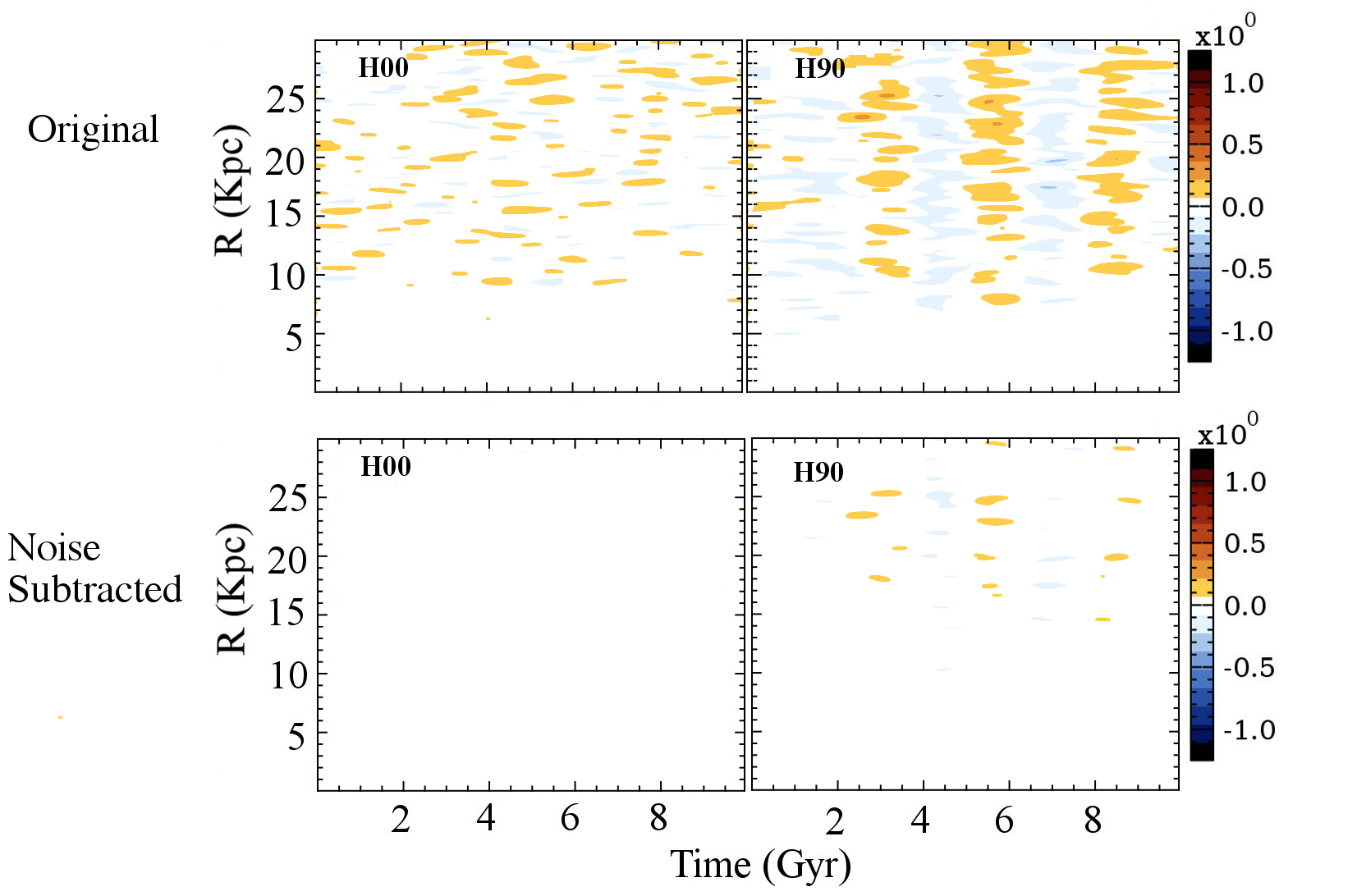}}
\caption{{\it The top row} shows the evolution of $\dot J$ within the inner $\ltorder 30$\,kpc of pure DM halos, as a function of cylindrical radius. Shown are models H00 with $\lambda = 0$, and H90 with $0.09$. No significant flow of $J$ is seen in these models, when compared to DM halos with embedded stellar disks in Figure\,\ref{fig:Jdotmap_sph} which uses the same color palette scale. {\it The bottom row} shows the subtraction of noise measured in the $\dot J$ transfer in both models (see text for details). After noise subtraction, the H00 halo no longer shows emission and absorption. H90 shows very low rate alternating emission and absorption of angular momentum between 15 to 30 kpc. No dynamical or secular evolution of density and angular momentum profiles has been detected.}
\label{fig:error}
\end{figure}

We have created a spherical nonrotating DM halo with isotropic velocity dispersion and the NFW density profile. The halo does not contain a disk and starts from equilibrium initial conditions. The model has been evolved for 10\,Gyr. We spin up this halo to $\lambda$ = $0.03$, $0.06$, $0.09, 0.1$, and $0.108$ using the same procedure from section\,\ref{sec:ICs}, and observe their subsequent evolution. Note that our spherical NFW halo reaches a maximum $\lambda=0.108$, when $100\%$ of particles are rotating in same direction. We call this sequence an H-sequence of diskless halos. 
 
The top frames of Figure\,\ref{fig:error} display the rate of $J$ flow inside the inner $\ltorder 30$\,kpc of the H00 and H90 models.  For $\lambda=0$ model, H00, the rate of $J$ flow, $\dot J$, exhibits noise only, as we show below. For models with increasing $\lambda$, a pattern develops of alternating, very weak emission and absorption of $J$ at radii $\gtorder 10$\,kpc. We have measured and found no growing low Fourier modes with $m=1-4$. No other changes pointing to internal evolution have been detected as well, such as in density and angular momentum profiles.

The bottom frames of Figure \ref{fig:error} show the noise subtracted plot. We calculated the variance in $\dot J$ over a 10\,Gyr and determined the width, $\sigma$, of the resulting Gaussian distribution of variance.  We then removed the $\dot J$ signal up to $3\sigma$ from the data in each figure, and plotted the resulting angular momentum rate of transfer.  The scale of the angular momentum transfer in the H90 halo now shows a very weak emission and absorption, basically corresponding to the minimal signal detectable in the color palette. This probably corresponds to the discreteness noise,
as $J$ is conserved within 0.1\% over 10\,Gyr.

Hence, all our models of diskless halos remain stable and do not form bars. Each halo maintains its original dispersion velocities and the original NFW density profile throughout the 10\,Gyr run. We include these results to show the stark contrast of halo evolution and angular momentum flow when the system hosts a stellar disk, e.g., Figure\,\ref{fig:Jdotmap_sph}. Hence we consider these diskless halos being stable both dynamically and secularly.
 
Our results do not contradict the models of \citet{dub94}, which obtained instability for a substantially oblate halo with axial ratio of $c/a=0.8$. In the NFW halos, most of the mass is located in the outer shells, and oblateness results in moving DM particles away from the rotation axis. As a result,
Kuijken \& Dubinski model had $\lambda=0.18$, almost a factor 2 larger in our halos with a maximal $\lambda$ of 0.1077. Hence, we have verified that DM halos within our $\lambda$ range are stable in the absence of stellar bars. 
 
We have run an additional test model that has our DM halo with $\lambda=0.09$ and a frozen disk potential. Such model with $\lambda=0$ was run by \citet{pete16} to examine the development of a DM bar without interaction from the stellar bar. However, in our test we push further into domain found by \citet{dub94} to be unstable, i.e., involving much higher spin. In this respect, our test is more challenging. We find no evidence of instability or bar formation in this halo as well. 

\subsection{Evolution of DM Bar Amplitude in Spinning Halos}
\label{sec:evo}

\begin{figure}
\centerline{\includegraphics[width=0.5\textwidth,angle=0] {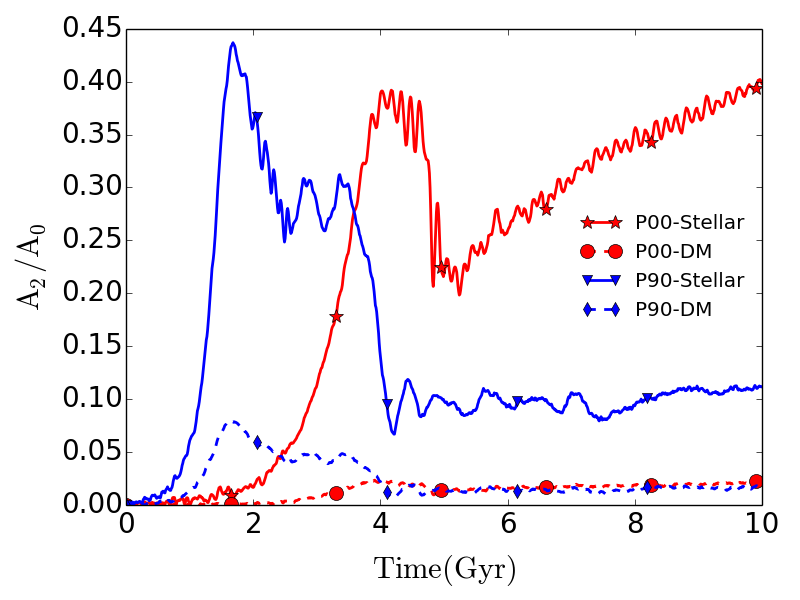}}
\caption{Evolution of stellar bar Fourier amplitudes, $A_2$, in the P00 and P90 models (solid lanes) and their corresponding DM bars (dashed lanes), normalized by the monopole term, $A_0$. These models represent the extremes of $\lambda$ sequence, i.e., $\lambda=0$ and 0.09. The limits of integration are described in the text.
}
\label{fig:comparison}
\end{figure}

\citet{coll18} have analyzed stellar bar evolution and the angular momentum redistribution in the disk of disk-halo systems with a range in $\lambda$. Here we focus on the response of DM halos under these conditions. Disks start axisymmetric, spontaneously break this symmetry and develop stellar bars --- a dynamical stage of evolution. These bars buckle vertically and either experience a secular growth or not, depending on the parent DM halo spin. To quantify this evolution in the DM, we measure departure of its density distribution from axial symmetry using Fourier mode amplitudes. The bar amplitude has been defined using the Fourier $m=2$ mode, namely,

\begin{eqnarray}
\frac{A_2}{A_0} = \frac{1}{A_0}\sum_{i=1}^{N_{\rm d}} m_{\rm i}\,e^{2i\phi_{\rm i}},
\end{eqnarray} 
although higher modes are not negligible, we ignore them here. The summation is performed over all disk particles with the mass $m=m_{\rm i}$ at azimuthal angles $\phi_{\rm i}$, for $R\leq 14$\,kpc, and $|z|\leq 5$\,kpc. The amplitude of the $m=2$ mode has been normalized by the monopole term $A_0$. The radial and vertical limits of summation correspond to the maximal length and well above the vertical thickness of stellar bars in P00 model.  
To measure amplitudes of DM bars, we followed the same procedure. For an unbiased comparison, we refrained from changing the radial and vertical limits of integration when measuring the amplitude of  the DM bars, although DM bars appear shorter and `fatter' than the stellar bars in all models.  

Figure\,\ref{fig:comparison} displays the evolution of the $A_2$ amplitudes for both stellar and DM bars, for two models at the extremes of the $\lambda$ sequence, namely $\lambda=0$ and 0.09, i.e., P00 and P90 models. The associated DM bars are much weaker than stellar bars, only reaching the maximum $A_2 \sim 0.02-0.08$ along the spin sequence, while the stellar bars reach $A_2 \cong 0.45$. Both phases of bar evolution, dynamical and secular, are highly dependent on the $\lambda$ of the parent halos, as seen in Figure\,\ref{fig:comparison}. Stellar bars in nonrotating and slowly rotating halos resume growth after buckling. Bar amplitude within faster spinning halos stagnates and shows no growth after buckling, during their secular evolution stage. 

\begin{figure}
\centerline{
\includegraphics[width=0.5\textwidth,angle=0] {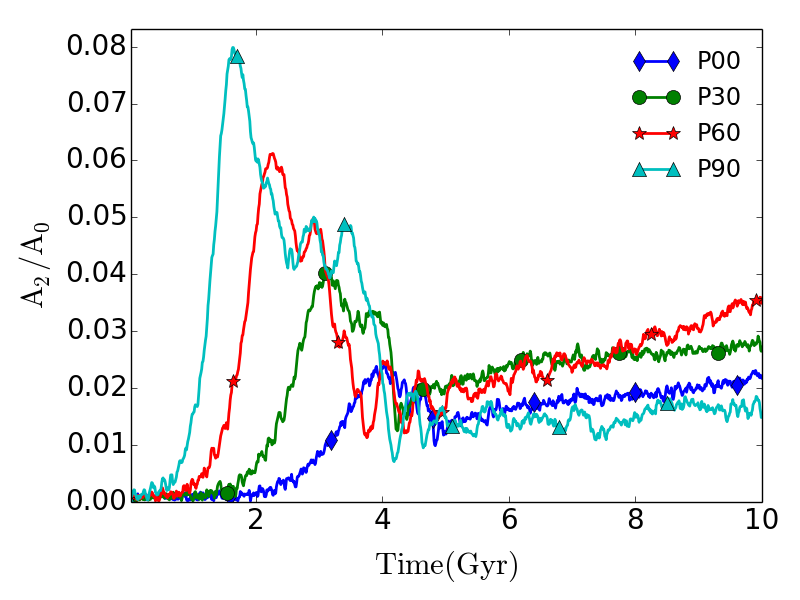}}
\caption{Fourier amplitude, $A_2$, normalized by the monopole term, $A_0$, evolution of the DM bars along the $\lambda$ sequence in our models. Note the strong dependency of DM bar maximal strength on $\lambda$ in the bar instability stage, and much weaker vertical spread in in $A_2$ in the secular evolution stage.}
\label{fig:a2lambda}
\end{figure}

\begin{figure}
\centerline{
 \includegraphics[width=0.5\textwidth,angle=0] {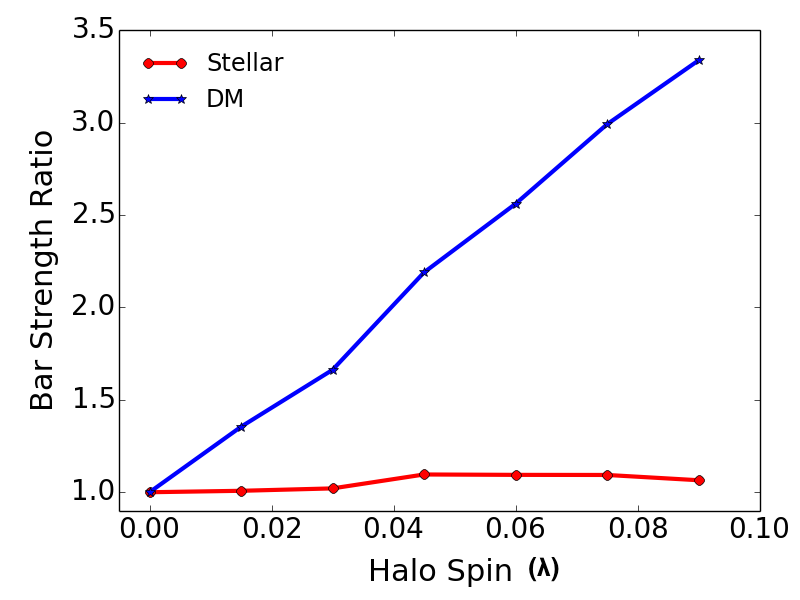}}
\caption{Comparison between stellar (red line) and DM (blue line) bars' maximal strength along the $\lambda$ sequence. Maximal bar amplitudes before buckling have been normalized by the maximal amplitudes of the stellar and DM bars, respectively, in the fiducial P00 model. Note that while the stellar bar amplitudes are largely independent of $\lambda$, the DM bar strength is strongly amplified by the parent halo spin.}
\label{fig:barratio}
\end{figure}

Figure\,\ref{fig:a2lambda} exhibits the evolutionary trends of stellar and DM bars along the $\lambda$ sequence. First, as $\lambda$ increases, the DM bar appears earlier, when compared to lower $\lambda$ models. Second, after buckling of stellar bars, DM bars inside large $\lambda$ halos do not regrow, similarly to their stellar counterparts, as noted first in \citet{coll18}. 

While the above evolution of DM bars is expected due to the evolution of the associated stellar bars, Figure\,\ref{fig:barratio} displays the idiosyncrasy in their behavior. Here we plot the maximal strength of DM and stellar bars, before buckling of stellar bars, normalized by the the maximal strength of the DM or stellar bar in the fiducial P00 model. The red line reflects the behavior of stellar bars in the Figure\,1 of \citet{coll18}, and stays flat. Meaning that the maximal pre-buckling amplitude of stellar bars is basically independent of $\lambda$. In contrast, the blue line representing the DM bars exhibits a dramatic increase with $\lambda$. For example, the P90 DM bar has its maximal pre-buckling strength amplified by a factor of $\sim 3.4$ compared to P00 model. 

We test whether this increase in the maximal amplitudes of DM bars along the halo spin sequence is related to the fraction of prograde populated DM orbits in our models. Table\,\ref{table:prograde} presents the halo spins and a fraction of prograde DM orbits. Comparing the maximal pre-buckling amplitudes of DM bars, we observe a direct dependency of the maximal $A_2$ on the fraction of prograde orbits, $f$, in the DM halo.  When $f = 1$, all DM particles are rotating in the same (prograde) direction, and when $f = 0.5$, the halo has $\lambda=0$.  

The amplification we observe in DM bar strength, along the $\lambda$ sequence, is not observed in the stellar bars, because most of the particles in the stellar disk are already on prograde orbits. To investigate the role of the prograde orbits along the halo spin sequence, we resort to the orbital spectral analysis in section\,\ref{sec:disk-halo}.

\begin{table}
\centering
\begin{tabular}{||c c c||} 
 \hline
 Model & $\lambda$ & $f$ \\ [0.5ex] 
 \hline
 P00 & 0.00 & 0.50 \\ 
 P30 & 0.03 & 0.62 \\
 P60 & 0.06 & 0.76 \\
 P90 & 0.09 & 0.88 \\ [1ex] 
 \hline
\end{tabular}
\caption{Fractions, $f$, of prograde DM orbits in our models.  }
\label{table:prograde}
\end{table}

\subsection{Spectral Orbital Analysis for Stellar Disks and DM Halos}
\label{sec:disk-halo}

\begin{figure*}
\centerline{
 \includegraphics[width=1.0\textwidth,angle=0] {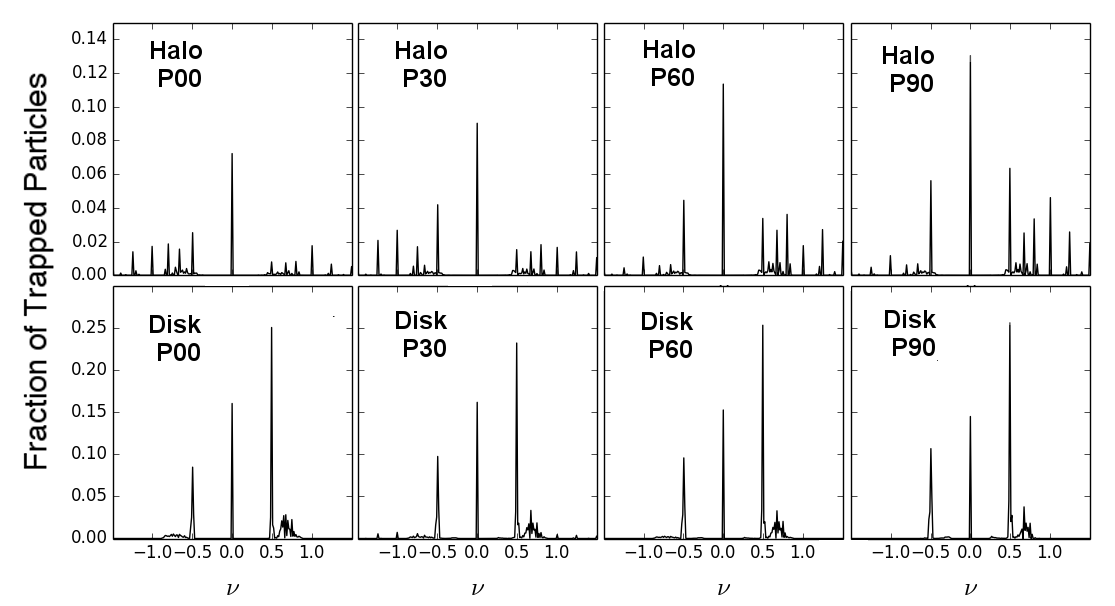}}
\caption{Comparison of the resonance trapping in DM halos and stellar disks along the $\lambda$ sequence. The $y$-axis shows the fraction of DM and stellar orbits trapped at each resonance, the ILR, CR, OLR and higher resonances. The $x$-axis gives the normalized frequency $\nu\equiv (\Omega-\Omega_{\rm b})/\kappa$ (see definitions in the text). The frequency bin is $\Delta\nu=0.01$. The main resonances are $\nu=0.5$ (the ILR), $\nu=0$ (the CR), and $\nu=-0.5$ (the OLR). The number of trapped particles is normalized by the total number of sampled orbits, i.e., 200,000. Shown are the frames for the halo (top frames) and disk (bottom frames) orbits, for four representative models, P00, P30, P60 and P90. The orbital spectral analysis has been performed before buckling, when {\it stellar} $A_2$ were near their maxima, and had approximately the same strength. The corresponding times are: $t=3.67$\,Gyr for P00, 2.82\,Gyr for P30, 2.01\,Gyr for P60, and 1.43\,Gyr for P90.
}
\label{fig:SA}
\end{figure*}

Next, we perform the orbital spectral analysis to determine the distribution of stellar and DM orbits with the normalized frequency $\nu$ for each of the models along the $\lambda$ sequence (section\,\ref{sec:spectral}). Our goal is to find the fraction of disk and halo orbits trapped at the main resonances, the ILR, CR, and the OLR, as well as at higher resonances. 

The orbital spectral analysis has been performed prior to buckling in each model, when stellar bar amplitudes have the same value. This allowed to measure the trapping efficiency at similar stages of bar evolution. Due to dependency of the bar strength on $\lambda$, similar $A_2$ occur at different times in each model \citep[][see also Figure\,\ref{fig:comparison}]{coll18}.  This comparison must be performed in the pre-buckling phase of evolution, because post-buckling disks within halos of $\lambda > 0.05$ essentially lack stellar bars, and only weak oval distortions are present, which cannot trap the orbits.

The resulting distribution of stellar and DM particles with $\nu$ is given in Figure\,\ref{fig:SA}. Both DM and stellar particles are concentrated at specific frequencies corresponding to the resonances. For disk stellar particles (the bottom frames), the main trapping corresponds to the ILR, $\nu=0.5$. Smaller fractions are trapped at the CR and the OLR, respectively. Previous work has clarified which resonances are mainly responsible for the angular momentum loss by the disk, and singled out the ILR as the main sponsor \citep{atha03,marti06,dub09}. Note that the trapping fraction of stellar orbits is independent of $\lambda$. 

A completely different picture emerges about the DM halo orbits trapped by the resonances (top frames). The CR resonance in the halo indeed remains the most efficient in trapping the DM orbits for all $\lambda$, as noted before for $\lambda=0$ models. But the fraction of trapped DM orbits by the CR depends on $\lambda$, increasing monotonically with the spin. This increase in the efficiency of trapping correlates nicely with the increase in the fraction of prograde orbits in the halo Table\,\ref{table:prograde}.

The ILR resonance traps small amount of DM particles, which has been noticed already in \citet{marti06}. In P00, the OLR is weak and the ILR is very weak. Other resonances are completely negligible. What is new here, is that the trapping ability of the ILR increases rapidly with $\lambda$, much faster than that of the CR.  For P90, the ILR is the second important resonance. Nearly the same effect occurs with the OLR. For P60 model, the OLR is barely more significant than the ILR, their roles have been reversed in P90, where the ILR dominates over the OLR. We also note the nonnegligible trapping by higher resonances for larger $\lambda$, especially for inner resonances with $\nu > 0.5$, but also for outer resonances with $\nu < -0.5$. When counted together, these resonances compete with the three main resonances in trapping efficiency of the DM orbits, especially the $\nu = 1$ resonance. 

To estimate the contribution to the angular momentum transfer by resonance trapped orbits, we use the orbital spectral analysis for two time snapshots. We measure the angular momentum lost by the stellar bar to the outer disk and to halo
during this time interval. The angular momentum lost due to resonance trapped DM orbits is obtained from difference in $J$ of these orbits at these two snapshots. Finally, we subtract the resonant angular momentum lost from the total $J$ lost --- this $J$ transfer is attributed to nonresonant interactions. The timing of two snapshots for the orbital spectral analysis is tuned to the moment the disks from the P00 and P90 models had lost the same amount of angular momentum. In the P00 model, we find that $\sim 50\%$ of the total angular momentum lost is due to resonant exchange. In the P90 model, this number is $\sim 88\%$. These numbers reproduce the values of $f$ (Table\,\ref{table:prograde}). This difference might seem small but it represents the angular momentum lost during a limited time, $\sim 1$\,Gyr, of a pre-buckling evolution. In the long run of 10\,Gyr, the P90 model disk loses much less $J$ than P00 disk, because the bar is essentially dissolved after buckling and the $J$ transfer is stopped. Overall, we observed a trend, in models with larger fraction of retrograde DM orbits, we find that the nonresonant $J$ transfer is more significant, as long as the stellar bar persists.

So, we conclude that the fraction of particles trapped at specific resonances is directly responsible for the angular momentum redistribution in the disk-halo system. Section \ref{sec:discussion} discusses the measured angular momentum transfer from orbital spectral analysis presented in figure \ref{fig:SA_DJ}.

For the first time we show that the efficiency of orbit trapping by the resonances in spinning halos depends on $\lambda$ and is directly proportional to the fraction of the prograde DM orbits. At the same time, the lower frames of Figure\,\ref{fig:SA} confirm that efficiency of trapping of the disk orbits is independent of $\lambda$, as expected. We observe a steady and monotonic increase of trapping efficiency by the halo resonances, the CR, ILR and OLR, as well as by higher resonances. 

\subsection{Sizes and Masses of DM Bars and Masses of Overall DM Response in Spinning Halos}
\label{sec:evo_size}

\citet{coll18} have calculated stellar bar sizes using two methods \citep{marti06}: measuring the maximal extent of an $x_1$ orbits from the characteristic diagram at each timestep, and by fitting ellipses to disk isodensity contours, obtaining the radial ellipticity profiles, $\epsilon(r)$, and determining where $\epsilon$ falls 15\% below its maximal value. For DM bar sizes, we apply the 2nd method in the DM halo equatorial slice.  
  
\begin{figure}
\centerline{
 \includegraphics[width=0.5\textwidth,angle=0] {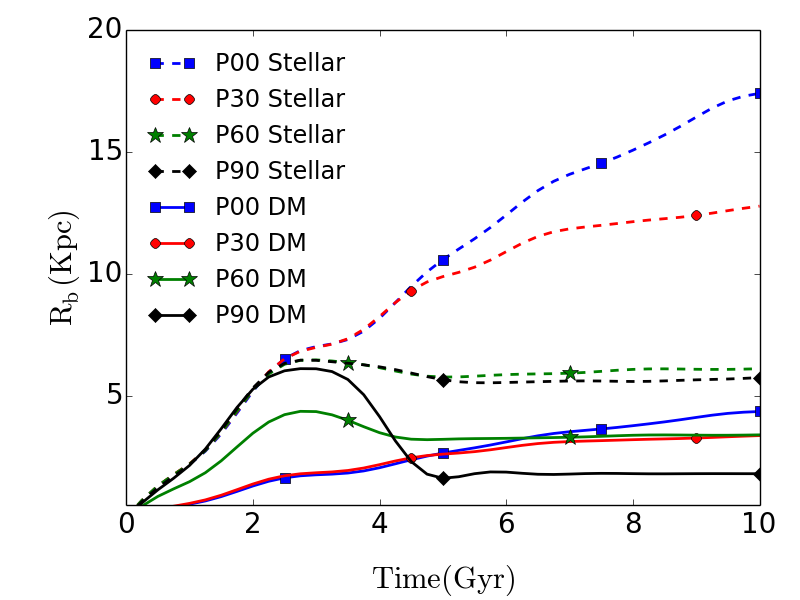}}
\caption{Evolution of DM (solid lines) and stellar (dashed lines)  bar sizes for models in the range of $\lambda\sim 0-0.09$. The stellar bar sizes have been determined from extension of $x_1$ orbits and measuring the bar ellipticity profile of isodensity contours in the $xy$-plane --- to the radius where ellipticity has decreased by 15\% from its maximum \citep{marti06,coll18}. The DM bar size have been obtained using the ellipticity profiles. 
}
\label{fig:barlength}
\end{figure}

Figure \ref{fig:barlength} shows the evolution stellar and DM bar sizes, $R_{\rm b}$. One observes substantial differences between these two components. For low $\lambda$ halos, the DM bars grow monotonically in size but remain a fraction of the corresponding stellar bars. For faster spinning halos, $\lambda$ \textgreater $0.03$, we observe that DM bar length rivals that of associated stellar bar. 

We observe three categories of $R_{\rm b}$ evolution --- growth, saturation and decline. All closely correlated with their stellar counterparts.
For $\lambda < 0.03$, the DM bar sizes grow monotonically. For the range of $\lambda\sim 0.03-0.05$, they saturate. And for the extreme, $\lambda\gtorder 0.06$, the DM bar sizes decline sharply after the buckling of stellar bars. This corresponds to the dissolution of stellar bars.

\begin{figure*}
\centerline{
 \includegraphics[width=1.0\textwidth,angle=0]{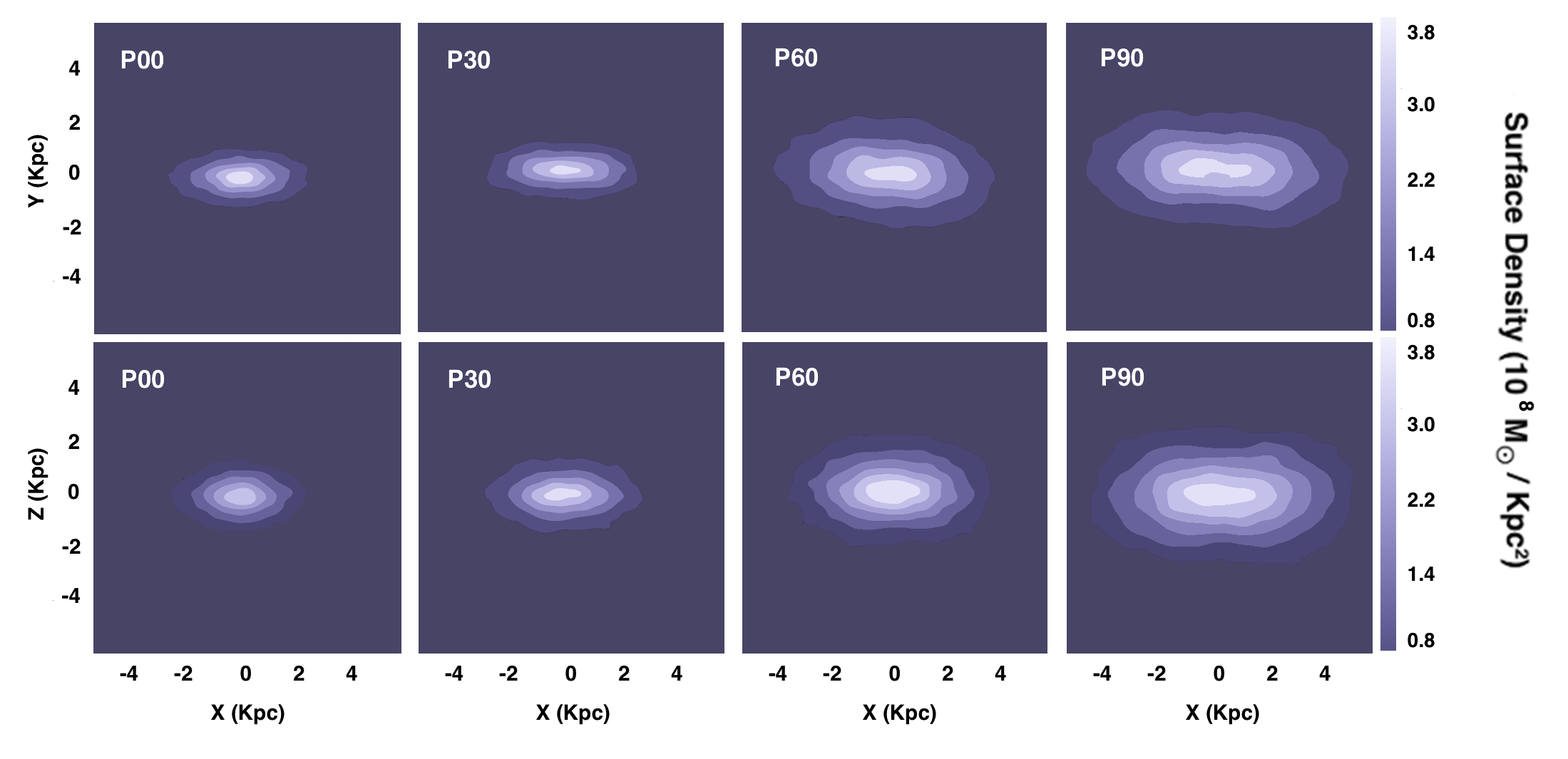}}
\caption{Comparison of face-on (top frames) and edge-on (bottom frames) DM bars along the $\lambda$ sequence. Shown are the surface density contours at the time near the maximal strength of these bars (times listed in Figure\,\ref{fig:SA}). The color palette provides the surface density in units of $10^8\,M_\odot\,{\rm kpc}^{-2}$. This figure displays only DM orbits trapped by inner resonances with $\nu > 0$, i.e., found inside the corotation radius, obtained from orbital spectral analysis in section\,\ref{sec:disk-halo} and from Figure\,5 in \citet{coll18}.   
}
\label{fig:Surface}
\end{figure*}

\begin{table}
\centering
\begin{tabular}{||c c c c c||} 
 \hline
  & P00 & P30 & P60 & P90 \\ [0.5ex] 
 \hline
DM Bar Mass & 0.30 & 0.52 & 0.81 & 1.05\\ 
Stellar Bar Mass &2.49& 2.49& 2.49& 2.49 \\
Ratio (DM/Stellar) & 0.12 & 0.21 & 0.33 & 0.42\\
  \hline
Mass of DM Response & & &  & \\ 
Outside the CR & 0.29 & 0.33 & 0.34 & 0.46\\
 \hline
\end{tabular}
\caption{Estimates of DM and stellar bar masses near their maximal strength in the pre-buckling stage, in units of $10^{10}\,M_\odot$. Also shown are the ratios of DM-to-stellar bar masses. The lower line displays the masses of resonant DM orbits outside the CR, in units of $10^{10}\,M_\odot$.}
\label{table:mass}
\end{table}

We now estimate DM and stellar bar masses. For the stellar bars, we adopt the major semi-axes and ellipticities from \citet{coll18}. We assume that the vertical thickness of a stellar bar is that of the disk and the peanut/boxy bulge. We count all the stellar masses within this volume.

For the DM bars, we use the above method to calculate their $R_{\rm b}$ in the $xy$-plane. The ellipticity of the isodensity contour crossing $R_{\rm b}$ in the equatorial plane provides us with the intermediate semi-axes of DM bars. Similarly, we find ellipticity of the isodensity contour which crosses the $R_{\rm b}$ point in the $xz$-plane, where $x$ axis is oriented along a stellar bar. This gives us the minor semi-axes of DM bars. As a next step, we calculate the volume of the ellipsoid using its axes, and measure the mass inside this figure. But, we do not expect all the DM orbits within the ellipsoid to be trapped by the stellar bar, thus forming the DM bar. 

\begin{figure}
\centerline{
 \includegraphics[width=0.5\textwidth,angle=0] {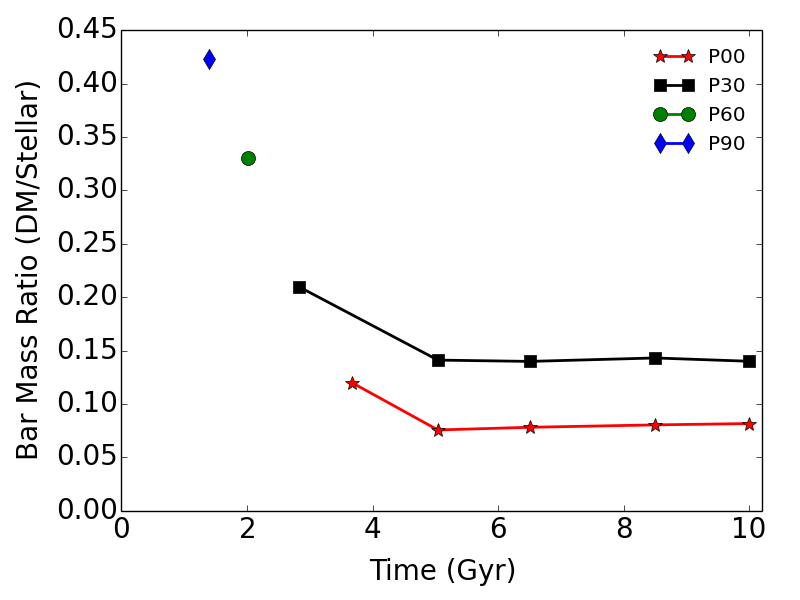}}
\caption{The DM-to-stellar bar mass ratio for models in the range of $\lambda=0-0.09$. The DM bar masses have been calculated from the orbital spectral analysis in section\,\ref{sec:disk-halo}, and include only DM orbits trapped inside the CR. The stellar bar masses have been calculated likewise. For all models, we provide mass ratios determined close to the maximal strength in the pre-buckling stage (the earliest points). For P00 and P30 additional 4 times each have been provided,
covering the evolution timescale up to 10\,Gyr. P60 and P90 stellar and DM bars essentially dissolved after buckling, and hence no orbital analysis has been performed.}
\label{fig:dmwithtime}
\end{figure}

The shapes of DM bars in our models are displayed in Figure\,\ref{fig:Surface}.  We plot the surface density of trapped DM particles inside the CR of the stellar bar found from spectral analysis in Figure\,\ref{fig:SA}. The top frames of the figure shows the face-on view of the DM bars. As $\lambda$ increases, the major and intermediate semi-axes of the DM bar also increase. The central region of the face-on P90 and P60 DM bars appear lopsided and dumbbell-shaped.

The bottom frames of Figure\,\ref{fig:Surface} shows the edge-on view of the DM bars in all models. As $\lambda$ increases, the number of orbits trapped at higher $|z|$ increases as well. The increase of the vertical extent of trapped DM region with $\lambda$ is in sharp contrast with the trapped stellar particles in the stellar bar, which is confined to the $z$-extent of the disk, which is geometrically thin. 

From orbital spectral analysis, we have obtained the percentage of trapped DM orbits by measuring the fraction of orbits inside the ellipsoid with calculated axes --- orbits that are in resonance with the stellar bar. We have {\it excluded} all the trapped orbits outside the CR. The resulting DM bar masses are then normalized by the associated stellar bar masses and given in Table\,\ref{table:mass}. Increasing $\lambda$ affects the DM bar sizes and masses substantially, by increasing the number of trapped orbits, while it has no effect or an adverse effect on the stellar bar masses. Therefore, in addition to the increased number of prograde orbits in the same volume, more orbits are trapped away from the equatorial plane, providing nonlinear amplification to the DM bar strength. Finally, we note that the DM response involves additional resonant orbits outside the CR. This contribution is shown in the last line of Table\,\ref{table:mass}.  

\begin{figure*}
\centerline{
 \includegraphics[width=1.0\textwidth,angle=0]{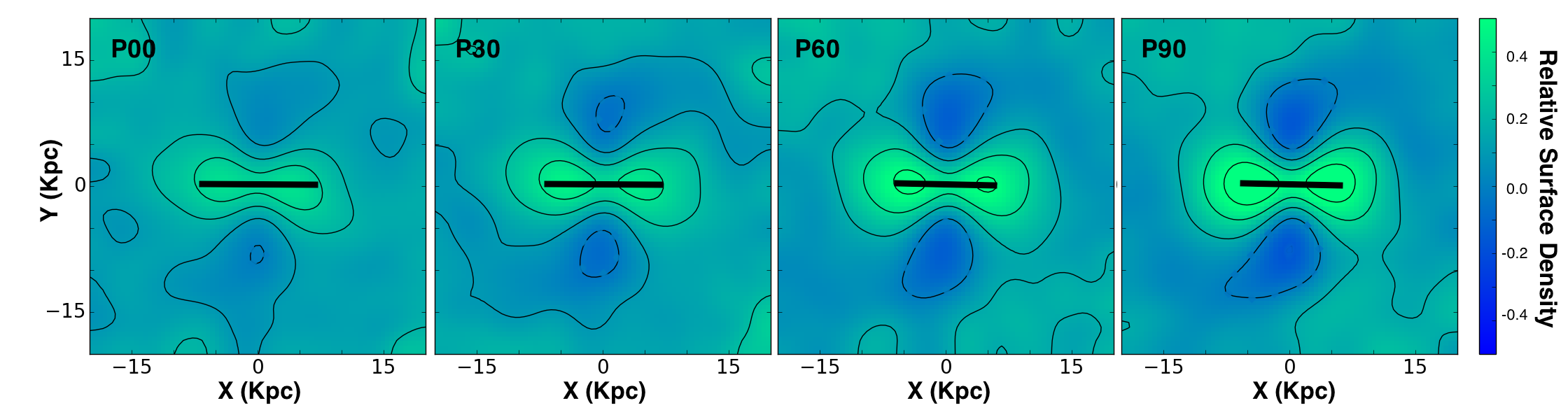}}
\caption{Surface density of the DM halos, at the times given in Figure\,\ref{fig:SA} caption, subtracting surface density at $t=0$, for all models. Surface densities were calculated in the $xy$-plane within a slice with $|z|\ltorder 3$\,kpc. Surface densities are given in the color palette. The position of stellar bar and its length are given by the straight line from \citet{coll18}. The positive contours are black solid lines and the negative ones are dashed lines. The outline of density enhancements and deficiencies delineate the DM bars and the DM gravitational wakes. Note that both extension and surface density perturbation amplitude of the DM response vary with increasing $\lambda$. 
}
\label{fig:wakes}
\end{figure*}

\begin{figure*}
\centerline{
 \includegraphics[width=1.0\textwidth,angle=0] {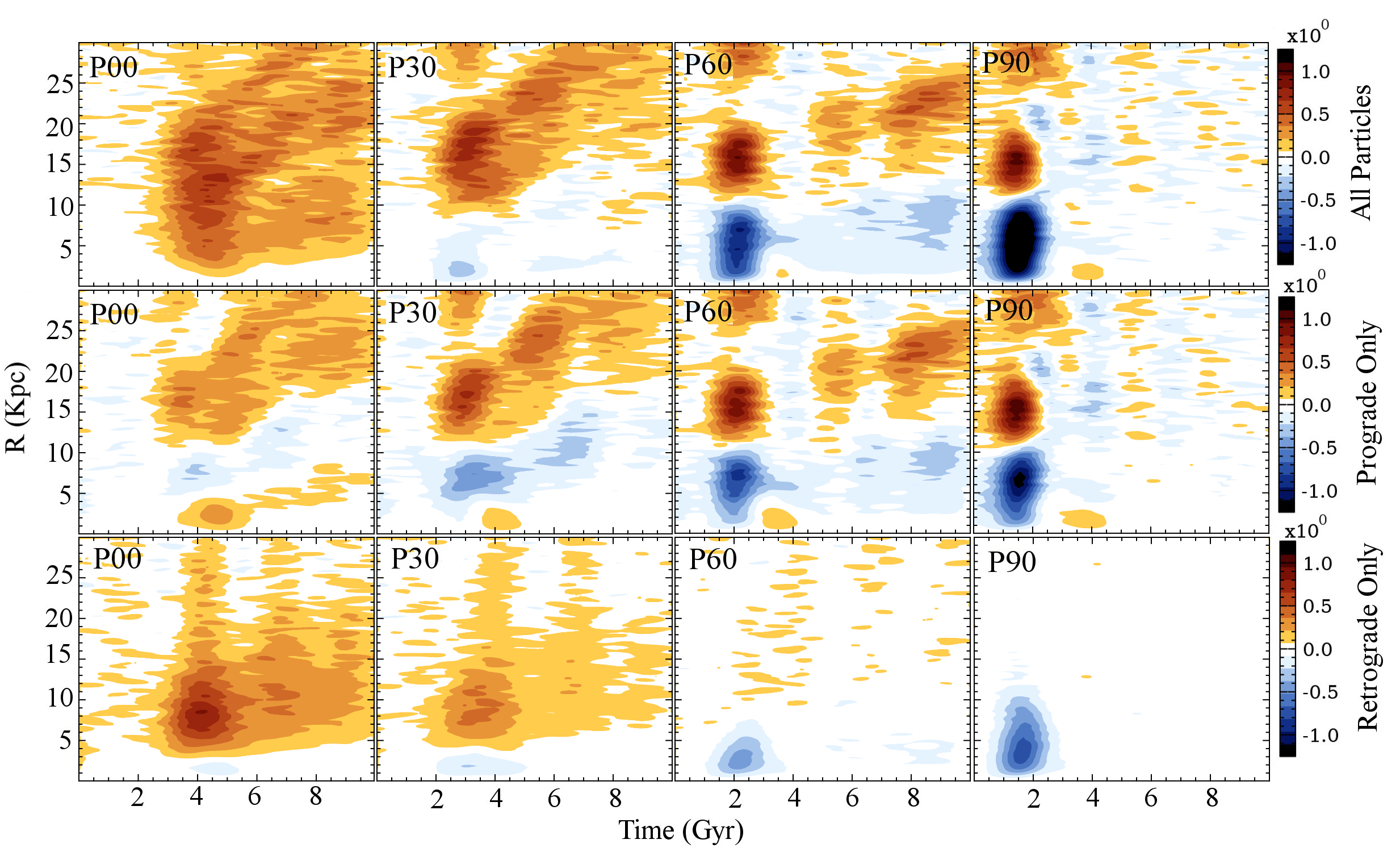}}
\caption{Rate of angular momentum, $\dot J$, emission and absorption by prograde and retrograde DM halo orbits as a function of a cylindrical radius $R$ and time, along the $\lambda$ sequence. The color palette corresponds to gain/loss rates in $J$ (i.e., red/blue),  using a logarithmic scale in color. The cylindrical shells binned at $\Delta R = 1$\,kpc and extend to $z=\pm 10$\,kpc. The top row includes both prograde and retrograde orbits in the DM halo. The middle row --- only the prograde orbits, and the bottom row --- only the retrograde orbits. The unit of angular momentum transfer rate used in the color palette is $10\,{\rm M_\odot\,kpc\,km\,s^{-1}\, yr^{-1}}$.
}
\label{fig:Jdotmap_sph}
\end{figure*}

From evolution of the obtained ratios of bar masses (Fig.\,\ref{fig:dmwithtime}), we note that in the low-$\lambda$ models, P00 and P30, the bar mass ratio drops abruptly with the stellar bar buckling, and continues to increase very slowly thereafter. For high-$\lambda$ models, P60 and P90, the bars basically dissolve, and no further measurements have been performed. Note that P30 resonances trap more DM orbits compared to P00 and hence its DM bar is more massive, and extends further out in $R$ and $z$, as is evident from figure\,\ref{fig:Surface}. In the highest spin halo, P90, the DM bar contributes more than 40\% of the stellar bar mass. Our error in the above estimates involves resonant DM orbits which lie outside the volume delineated by $R=29$\,kpc and $|z|=10$\,kpc --- altogether $\sim 3\%$ of resonant DM orbits. For the edge on estimation we look at the limit of $|y|=10$ which in the $xy$-plane.

The above trend in the DM bars can be extended to the overall DM response to perturbation by the underlying stellar bar. Figure\,\ref{fig:wakes} displays the full response, which includes the DM bars and associated gravitational wakes which extend from inside the corotation radius to the OLR. Both the extent and the amplitude of the response are growing with the halo spin.

\subsection{Angular Momentum Transfer for Prograde and Retrograde DM Orbits}
\label{sec:JpureDM}

The DM halo of the P00 model consists of equal fractions of prograde and retrograde DM orbits, with respect to the disk rotation. These fractions change when moving along the $\lambda$ sequence, as shown in Table\,\ref{table:prograde}. As a next step, we analyze contributions of prograde and retrograde DM orbits to the rate of angular momentum transfer along the $\lambda$ sequence.

To calculate the angular momentum flow in the system, we use the method prescribed in \citet{villa09} and also used in \citet{long14} and \citet{coll18}. We bin the halo into cylindrical shells of $\Delta R = 1$\,kpc, parallel to its rotation axis, and create a two-dimensional map of the rate of change of $J$, i.e., ${\dot J}$, in each shell as a function of $R$ and time. The maps are color coded to show positive transfer of $J$ in red and negative transfer of $J$ in blue. The color palette has been normalized the same way for each figure. This method follows the total angular momentum rate of transfer between the disk and the DM halo, though here we show only the halo. As we discuss in section\,\ref{sec:discussion}, the $J$ transfer is not only limited to disk-to-halo, but the DM halo also `talks' to itself.  The disk has been displayed in \citet{coll18}.

As a first step, we calculate $\dot J$ --- the rate of the angular momentum transfer to, away, and within the DM halo for models along the $\lambda$ sequence (Fig.\,\ref{fig:Jdotmap_sph}, top frames). These frames have been shown already in \citet{coll18} in lower resolution. The P00 model halo exhibits a pure absorption of $J$ from the disk. Three resonances appear prominent in its frame --- the ILR, CR and OLR. They move out with time due to the stellar bar slowdown.  These resonances can be traced as well after buckling. Most of the transfer happens close to the time when the $A_2$ of the stellar bar is at the maximum of $A_2$ (before and after buckling).   In this model of a nonrotating halo, the stellar bar strength recovers after buckling and reaches its pre-buckling maximum at $t\sim 10$\,Gyr (e.g., Fig.\,\ref{fig:comparison}). 

In the P30 model, prominent changes occur. The halo ILR region shows no emission but a weak absorption of $J$. The $J$ transfer rate in the OLR region is enhanced and the CR rate stays unchanged and weakens thereafter. In this model, the stellar bar although recovers part of its original strength after buckling, it falls short of the maximal $A_2$ in the pre-buckling stage.

The P60 model shows a dramatic difference when compared with the lower $\lambda$ models. The CR dominates in absorption, while the ILR is now very prominent in emission. This trend continues to the P90 model. In these two models, the stellar bars do not grow after the buckling. In fact, the loss of strength in buckling is substantial, and the bars dissolve into weak oval distortions.

By separating the angular momentum flow for prograde and retrograde DM orbits, we gain some insight into $J$ redistribution in the system. The middle row in Figure\,\ref{fig:Jdotmap_sph} displays the $\dot J$ along the $\lambda$ sequence for prograde orbits only. For the P00, separation into prograde and retrograde orbits underlines the diminishing role of the ILR for former orbits compared to the upper row of all orbits, and appearance of a very weak absorption there. On the other hand, the retrograde orbits contribute massively to $\dot J$ in the ILR region. 

With increasing $\lambda$ we observe an increasing emission by the ILR for prograde orbits, decreasing absorption and switching from emission to absorption, with an overall decrease of importance of $J$ flow in the ILR region. This can be explained by a fractional decrease of the retrograde orbits along the $\lambda$ sequence.  When comparing Figure\,\ref{fig:Jdotmap_sph} to Figure\,\ref{fig:SA} we see that the nonnegligible increase in halo emission of $J$ coincides with the increase in resonant orbits within the ILR, especially at $\nu=1$.

\section{Discussion}
\label{sec:discussion}

Using high-resolution numerical simulations, we analyze evolution of the DM bars in disk-halo systems over 10\,Gyrs, which form in response to the stellar bars in the embedded disks. We focus on spinning DM halos with cosmological spin $\lambda\sim 0 - 0.09$, and investigate how $\lambda$ affects the evolution of DM bars, their morphology, strength, mass, size, the flow of angular momentum in these systems, and implications for the stellar disk evolution. For this purpose, we use a representative model for DM mass and angular momentum distributions, leaving consideration of a wide range of these to future work.

We start with outlining our main results.
First, we find that the maximal strength of induced DM bars depends strongly on the halo spin, while the maximal strength of stellar bars is indifferent to $\lambda$. Hence, the maximal strength of DM bars depends on the fraction of prograde DM orbits in the host halos. Second, the efficiency of trapping DM halo orbits by the resonances, including the ILR, OLR and CR, as well as higher resonances, depends on $\lambda$. We show this explicitly by means of the orbital spectral analysis. It remains important to show that increase in the trapping efficiency results in increase in the angular momentum transfer by the resonances, and we address this issue below. 

Third, higher resonances, inside the ILR and outside the OLR, become progressively more important in trapping the DM orbits and hence for angular momentum flow in the system with $\lambda$. We quantify this transfer below. The DM halos not only receive their $J$ from the disks, but actively transfer it to larger radii by means of these resonances. In this respect, the halo `talks' not only to the underlying stellar disk but to itself as well.   

Fourth, we analyzed the roles of prograde and retrograde DM orbits contribution to $\dot J$. The prograde orbits dominate absorption of $J$ at CR for all $\lambda$,  and show an increasing emission of $J$ with $\lambda$. On the other hand, the retrograde orbits absorb at low $\lambda$, and emit $J$ at higher $\lambda$.

The robustness of stellar bars was called into question when halo spin has been introduced \citep{long14,coll18}. The halo spin $\lambda$ has a dramatic effect on the dynamical and secular evolution of the stellar bar, and this holds true for DM bars as well. Because the DM bars represent the halo response to the stellar bars perturbation, it is not surprising that they mimic evolution of stellar bars to a larger extent. What is surprising is that the DM response is so nonlinear --- one cannot predict the amplitude of a DM bar by simply scaling it with the stellar bar. Its behavior is more complex than this. 

The maximal amplitude of stellar response in galactic disks appears to be completely indifferent in its dynamical stage (i.e., pre-buckling) to the host halo spin. This is in a sharp contrast with the DM response --- in the range of $\lambda=0-0.09$, the amplitude of DM bar varies by factor of $\sim 3.5$.  What is the reason for such a sharp difference between the stellar and DM responses? 

Plausibly, the answer lies in the increase of the number of prograde DM orbits, but this increase is less than a factor of two, as shown by Table\,\ref{table:prograde}. Note that even in $\lambda=0$ halo, 50\% of the DM orbits are prograde. Hence additional cause must be at play. A cause which is absent in the case of a stellar disk. 

Orbits in a stellar disk prior to bar instability are largely prograde, and the geometrical thickness of the disk naturally limits their $z$-extent. On the other hand, even for the $\lambda=0.09$ spherical halo, when almost 90\% of DM orbits are prograde, the vertical extent of the DM orbital trapping is not strictly limited, and higher latitude orbits can be affected and trapped, for example by orbits closer to the equatorial plane coupling to the higher $z$ orbits. In this case, we should observe that both the mass and size of a DM bar grow with the spin. This is exactly what is shown in Figure\,\ref{fig:Surface} and Table\,\ref{table:prograde}. Moreover, this can be seen directly in Figure\,\ref{fig:Jdotmap_sph}, where loss of angular momentum in P60 and P90 models can be observed for DM orbits lying close to the equatorial plane. This $J$ is absorbed by the higher altitude DM orbits.  

We conclude that the main difference in DM response compared to the stellar response lies in the availability of orbits capable of resonating with the perturber and being trapped by numerous resonances. The orbital spectral analysis provides the necessary insight into properties of DM bars, their evolution and the associated intricacies of angular momentum transfer within the disk-halo system. Figure\,\ref{fig:SA} exhibits monotonous increase in the trapping efficiency with $\lambda$ of the three main resonances and additional higher resonances up to $|\nu|\sim 1.5$ searched by us. This increase can be observed already in P30 compared to P00, and is supported by similar frames in Figure\,\ref{fig:Surface}. For higher $\lambda$, the change is much more dramatic. 

This progression in DM response with $\lambda$ was not observed by \citet{pete16}, because they limited their models by $\lambda=0.03$, and hence passed over this effect. We have tested all our halos without and with {\it frozen} stellar disk potential, and found that all halos are stable over 10\,Gyr, and that previous claims in the literature about developing instabilities in these systems are not substantiated by our analysis, which extends to $\lambda\sim 0.108$ --- the maximal spin attainable in spherical NFW halos (section\,\ref{sec:pureDM}).  

We have compared the properties of DM bars, e.g., their masses, for P00 and P30 models, with  \citet{pete16}, and find a very good agreement. Specifically, the DM-to-stellar bar mass ratio is $\sim 0.1$ for P00. We find, however, that the DM response goes well beyond the DM bar, and additional DM mass participates in the gravitational wake, as shown in Table\,\ref{table:mass}.
For $\lambda\leq 0.03$ models, we do observe a very slow growth for the mass ratio of the bars, after the buckling, again in agreement with Peterson et al. Yet during the buckling instability of stellar bars, we find that this ratio drops substantially. For higher $\lambda$, the secular growth is completely suppressed and  both types of bars essentially dissolve during buckling. 

In addition to varying in length and mass, we have measured the angular separation between the stellar and DM bars. With stellar bar leading, this angle decreases with $\lambda$, if measured for same strength of stellar bars. This is another indication that DM bars become stronger with $\lambda$ (Figure\,\ref{fig:a2lambda}).

Next, we deal with the efficiency of $J$ transfer from the stellar disk to the DM halo by orbits trapped at the resonances. Specifically, we ask what is the fraction of transferred $J$, attributed to the action of the resonances, compared to the total amount of $J$ transferred over a fixed time interval. To answer this question, we choose two moments of time, $t_1$ and $t_2$ for each model, P00 and P90, when stellar disks lose identical amounts of angular momentum. In other words, we choose these times in such a way that each of the stellar disks have lost the same amount, $\Delta J$(P00)$=\Delta J$(P90). Because the total $J$ should be conserved, the DM halos in both models should absorb the same amount of the angular momentum. 

\begin{figure}
\centerline{
 \includegraphics[width=.5\textwidth,angle=0] {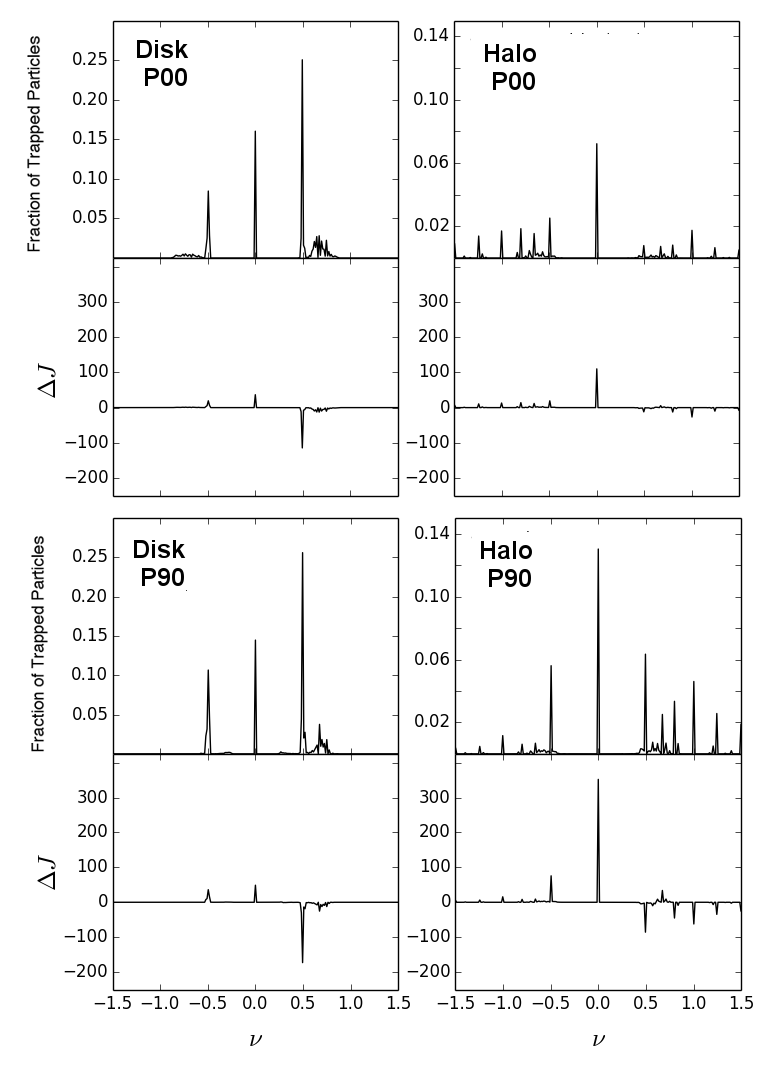}}
\caption{{\it Top:} P00 model --- fractions of trapped stellar orbits and the change of angular momentum, $\Delta J$, by these orbits (left frame), and fractions of trapped DM orbits and their $\Delta J$ (right frame). {\it Bottom:} P90 model --- same as P00. Trapping has been calculated for the three main resonances (ILR, CR, OLR) and for higher resonances, up to $|\nu|=1.5$ (see Figure\,\ref{fig:SA} for more details). The $\Delta J$ units are $10^{10}\,{\rm M_\odot\,kpc\,km\,s^{-1}}$. Note, the halo and disk $y$-axes are scaled differently for clarity. Note, the stellar bars exhibit the same trapping ability and lose identical $\Delta J$ in both models. Hence, both P00 and P90 halos gain the same $\Delta J$, with a notable difference: the resonant $J$ transfer dominates over the nonresonant one in P90, while not in P00 (see text for more details). 
}
\label{fig:SA_DJ}
\end{figure}

We calculate the amount of $J_{\rm res}$ absorbed by the halo resonances in each model during this time interval, using our method of orbit spectral analysis. Then find the ratio $J_{\rm res}/\Delta J$, for each model and compare them. This will provide a rough estimate of the efficiency of $J$ transfer by the resonances as a function of the halo spin $\lambda$. 

Note that \citet{lynd72} have demonstrated that $J$-transfer happens at the resonances. But they did not rule out some contribution from non-resonant torques to this process. On the other, orbits that are non-resonant at some time could resonate at some other time, making the careful check a difficult task. We limit our conclusions to specific times in the evolution of the system only, leaving the more general conclusions outside the scope of the present work. 

Figure\,\ref{fig:SA_DJ} shows the results of our analysis, for P00 and P90 models, and includes the orbital trapping in the stellar disks and DM halos, as well as calculation of angular momentum change at each resonance, up to $|\nu|=1.5$, between $t_1$ and $t_2$. This figure confirms that the {\it stellar} orbital trapping is independent of $\lambda$, i.e., it is the same in P00 and P90, even comparing individual resonances. The disk lost most of the angular momentum by the ILR, and absorbed a small amounts of angular momentum by the CR and the OLR. 

The halo action is very different. The main resonance absorbing $J$ is the CR, but in P90 it absorbs by far more angular momentum than in P00. Where does the excess of $J$ come from? We note that all the other resonances become much more active in P90 halo compared to P00. Some absorb and some emit $J$, but net $J$ is absorbed. In fact, only resonances inside the CR lose $J$, while the outside CR resonances absorb it. We return to this point below, and see it as a proof that the inner DM halo exchanges $J$ with the DM further out --- the halo `talks' to itself.

We now compare the numbers obtained from orbital analysis of Figure\,\ref{fig:SA_DJ}. We define the angular momentum unit as $10^{10}\,{\rm M_\odot\,kpc\,km\,s^{-1}}$, and all angular momentum values in the following are given in these units. The total $J$ lost by P00 and P90 disks, $\Delta J\approx 278$, is the same by construction. The net resonant transfer using $\Delta J_{\rm res}$ in P00 halo is $\approx 138$, and in P90 is 246. The net transfer by nonresonant torques in P00 and P90 respectively is about 139 and 31. We can estimate the fraction of $\Delta J$ transferred by the resonances, $\sim 50\%$ in P00, and 88\% in P90. Hence, the efficiency of {\it resonant} $J$ transfer is scaling directly with the fraction of prograde orbits in the DM halo, $f$, and therefore with its spin $\lambda$. 

An additional interesting point about the angular momentum transfer by the resonances can be observed in the P90 halo frame of Figure\,\ref{fig:SA_DJ}. This halo obtained about a net of $\Delta J_{\rm res}\sim 246$ in resonant transfer. This net angular momentum is made of absorbed $\sim 589$ and emitted $\sim 341$. Where is this excess of absorption coming from? It cannot come from the disk, as this number is quoted above and is smaller. However,  Figure\,\ref{fig:SA_DJ} reveals that the excess absorption by the CR and the OLR originates from the emission by the ILR and higher resonances inside the CR. The puzzle is resolved by  Figure\,\ref{fig:Jdotmap_sph} ---  this emission by the inner resonances comes from the inner halo, $<10$\,kpc, while absorption is performed by the region of DM halo at larger radii, $\sim 10-30$\,kpc. This is what we meant above by  the DM halo in faster spinning halos `talks' to itself.

We have discussed the rates of the $J$ transfer along the $\lambda$ sequence in section\,\ref{sec:JpureDM}. Here, we take a closer look at these rates, focusing separately on prograde and retrograde orbits in the DM halos. The two bottom rows of Fig.\,\ref{fig:Jdotmap_sph} show the angular momentum transfer by prograde and retrograde DM orbits, respectively. Moving along the $\lambda$ sequence, in the P00 model, the retrograde orbits primarily gain angular momentum, and doing this more intensely than the prograde ones. When $\lambda$ increases, the prograde orbits switch to losing $J$ within the inner 10\,kpc. This change in $\dot J$ is accompanied with an increase of absorption by the DM halo orbits at larger radii. At the same time, the retrograde orbits play a lesser role in the absorption, and switch to emission inside the inner 10\,kpc. Moving along $\lambda$, there is a smaller number of retrograde orbits at $t=0$, by construction. So, the P90 model displays only emission in the inner region, before the stellar DM bars dissolve following buckling. 
 
 \begin{figure*}
\centerline{
 \includegraphics[width=1.\textwidth,angle=0] {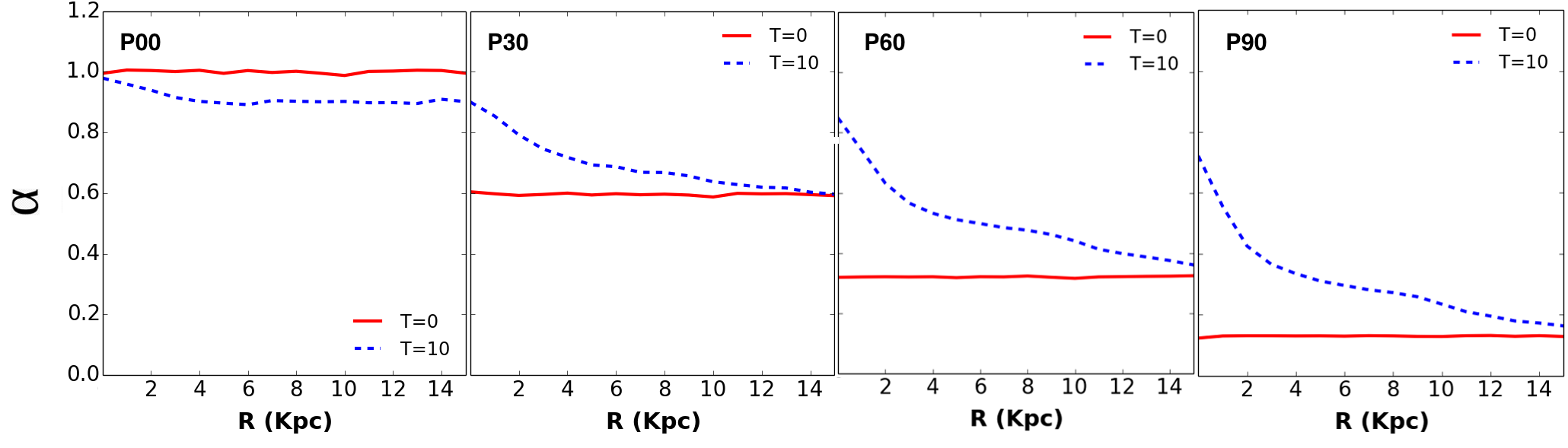}}
\caption{Ratio of retrograde particle number to prograde particle number, $\alpha$, for the inner halo, $R <  30$\,kpc, at two timesteps in the DM halo, at $t=0$ (solid line) and $t=10$\,Gyr (dotted line), for different models with $\lambda$ increasing to the right.}
\label{fig:rho}
\end{figure*}
 
The total number of DM and stellar particles is conserved in these simulations, but what about the ratio of retrograde to prograde DM orbits, $\alpha$? Figure\,\ref{fig:rho} shows this ratio for two models with extreme halo spins, P00 and P90, and doing this at $t=0$ and $t=10$\,Gyr. 

Clearly, $\alpha$ evolves with time, and in a non-uniform manner. At $t=0$, $\alpha$ is constant with radius for all models. For P00, where  $\alpha=1$ for $t=0$, at the end of the simulation, $\alpha\sim 0.9$ and is still relatively constant with radius. but for higher $\lambda$ models, at later time, $\alpha$ develops a profile in $r$, and increases substantially at smaller radii for higher $\lambda$ models. For example, in P90, as the stellar bar (and simultaneously the DM bar) develops and gains strength, the number of prograde orbits decreases within the stellar disk radius, more so in the central region, which becomes prograde. The prograde orbits become retrograde when they are trapped by the bar. The DM bar emits $J$ as these retrograde particles become trapped. We observe this loss in Figure\,\ref{fig:Jdotmap_sph}. Within the region of central $\sim 10$\,kpc. The ratio $\alpha$ has increased from $\sim 0.12$ at the start of the simulation, to $\sim 0.7$. 

One can ask whether the 'depth' of the buckling instability, as measured by the stellar $A_2$, depends on the number of retrograde orbits in the halo. Our initial conditions have identical disks embedded in DM halos of a differing spin. Analyzing the resonance trapping by the stellar disk at similar bar strengths inside the spinning halos (section\,\ref{sec:disk-halo}), we found that the trapping efficiency of stellar orbits correlates strongly with $\lambda$. We also measured $A_2$ and found that pre-buckling bar strength does not depend on the halo spin. Why then does the loss of strength by the stellar bar during buckling vary so much with $\lambda$? 

Figure\,\ref{fig:rho} provides some, yet not fully compelling explanation for this effect. Substantial loss of strength by the stellar bar during the buckling instability is accompanied by equal loss of strength by the associated DM bar. The DM orbits de-correlate their orientation during this short time period. This is expected to lead to a sharp reduction in the efficiency of resonance trapping. As shown in \citet{coll18}, the disk is heated up by the appearance of a large number of low angular momentum orbits released by the bar. DM bars follow this trend and release the DM orbits. A larger fraction of these orbits are retrograde for large $\lambda$, and most of them are found in the stellar bar region, as shown in this figure. Because DM bars strength depends on $\lambda$, and increases with it, this de-correlation will have a stronger effect on higher spin halos compared to low spin ones, and on the underlying stellar disks.   

We note, that models with strong stellar bars are expected to be accompanied by a strong DM bar component.  This is true especially for halos with larger spin, but {\it only} for $\lambda > 0$ ! If the halo spin can be measured, a strong case can be made for aiming direct detection DM experiments at these spinning halos with strong stellar bars, as we expect a large volume and DM mass to be trapped and accompanying the stellar bar.

Our choice of $J$ distribution with radius is not unique but is severely constrained by the NFW density profile and the halo shape. Within these limits, one can imagine the following $J$ distribution --- the inner halos remains as in P00 model, while the outer halo has the fraction of prograde particles modified, i.e., $f\sim 0.88$, as in P90 model. How does this affect the stellar bar evolution? 

To test this, we have analyzed the rate of $J$ transfer in our models, and especially the spatial extent of $\dot J$ map in Figure\,\ref{fig:Jdotmap_sph}.  We find that the part of the DM halo that is active in $J$ redistribution extends to the box of $R\sim 30$\,kpc and $z\sim 10$\,kpc, few time the stellar bar size. The stellar disk communicates with the halo within this volume, which includes the volume in which the halo `talks' to itself. The angular momentum outside this box has no effect on the interior and so on the stellar disk. Yet, one should exercise caution in this respect.

If $J$ is injected from outside into the halo surrounding the above box, it can and will propagate inwards on a secular timescale. As this process is irreversible, it reminds us of a `diffusion.' Strictly speaking, it cannot be applied to a collisionless system, but was nevertheless argued by \citet{lynd67}, by introducing 
coarse-grained and fine-grained distribution functions. We have tested this by spinning up the P00 halo to $f\sim 0.88$, but {\it only} outside 40\,kpc, and observed the $J$ slow propagation inwards.

In summary, the total DM response to the stellar bar involves mass of the order of the stellar mass, as follows from 
Table\,\ref{table:mass}.

\section{Conclusions}
\label{sec:conc}

We present results of high resolution numerical simulations of stellar disks embedded in spherical DM halos, with a range of cosmological spin parameter, $\lambda\sim 0-0.09$, and a representative angular momentum distribution. In this work we focus on the DM response to the developing stellar bars in the underlying galactic disks. This DM halo response evolves as a DM bar, and we investigate its role in the angular momentum transfer in the disk-halo systems. To address an ambiguity regarding global stability for spinning spherical halos, we have also tested models of diskless spinning DM halos, for $\lambda\sim 0-0.108$. The upper limit of $\lambda$ follows from a model with all DM particles on prograde orbits. Moreover, we have tested the model for DM halo with $\lambda=0.09$ and a frozen stellar disk potential. 
 
Our main conclusions are as follows:

\begin{enumerate}
\item We have shown that diskless DM spinning halos are stable to bar instabilities, and maintain their shape and velocity distributions. We have tested a range of halos, $\lambda = 0.00$ to $0.108$ and found that all our DM halos are stable in the absence of embedded stellar disks. We put to rest the idea that these halos can be globally unstable and develop the $m=2$ Fourier mode. DM halos with $\lambda=0.09$ remained stable even in the presence of an embedded {\it frozen} disk potential. Hence all the DM bars obtained in our simulations have been triggered by the developing stellar bars, i.e., have been induced by them.
  
\item The strength of DM bars depends strongly on the DM halo spin, and, therefore, on the fraction of prograde orbits\footnote{This statement should be taken with additional constraints imposed by the halo shape and the NFW profile, as discussed in section\,\ref{sec:discussion}.} within a volume substantially larger than the stellar disk radius. The maximal Fourier amplitude of the DM bars increases by a factor of $\sim 3.4$ when $\lambda$ increases from 0 to 0.09. This is in sharp contrast with stellar bars whose pre-buckling amplitude is independent of $\lambda$. For a disk containing about 98\% of its mass within 17\,kpc, the DM volume affected lies within a sphere with $\sim 30$\,kpc radius.

\item The efficiency of resonance trapping of DM orbits by the stellar bar increases with $\lambda$. This includes the main resonances, the ILR, CR and the OLR, as well as higher resonances, both inside and outside the CR. This was shown by invoking orbital spectral analysis. 

\item The angular momentum transfer from stellar disks to DM parent halos is maintained by both resonant and nonresonant orbits. But the efficiency of angular momentum transfer by the resonances increases with $\lambda$, from about 50\% at $\lambda=0$ to $\sim 88\%$ for $\lambda=0.09$. Furthermore, different resonances exchange the angular momentum with increasing $\lambda$. For example, the ILR emits $J$ while the CR and the OLR absorb it. In other words, at higher $\lambda$, the halo `talks' to itself by means of exchanging angular momentum, which flows from the inner resonances, e.g., the ILR, to larger radii, e.g., the CR and the OLR. Higher resonances become more important with an increase of halo spin..

\item Prograde and retrograde DM orbits play different roles in the angular momentum transfer in disk-halo systems. Prograde orbits dominate absorption of angular momentum at the CR for all $\lambda$, while retrograde orbits absorb at low $\lambda$ and emitting the angular momentum at higher $\lambda$. The fraction of retrograde DM orbits increases with time during the bar instability, compared to the initial conditions, where the fraction of retrograde orbits is constant with radius, by construction. This increase is more profound with increasing $\lambda$.

\item We find that the mass, length, and shape of DM bars have a strong dependence on the parent halo spin. The most massive, long and strong DM bars are found in spinning halos before buckling. 

\item The existence of the DM bar requires a stellar bar. The angle between the stellar and DM bars decreases with an increasing halo spin. The DM bars lag behind the stellar bars. 

\item The overall DM response to the underlying stellar bar involves mass of the order of the stellar bar mass. Furthermore, the active part of the DM halo which participates in the angular momentum redistribution in the system constitute a volume of $R\ltorder 30$\,kpc and $|z|\ltorder 10$\,kpc.

\end{enumerate}

Stellar bars have been studied numerically for about half a century. Yet they still pose unanswered questions. To a large extent they are the prime internal factor which shapes disk galaxy evolution. What makes them even more interesting is that they provide a strong link to host DM halos. Study of angular momentum exchange between the disk and halo components can shed new light on the evolution of galaxies which is driven by competition between internal and external factors.  The current work will be followed by a careful study of the effect of DM mass and angular momentum distributions on the bar evolution.

Cosmological simulations will approach the minimal resolution required to study interactions in disk-halo systems in the near future. Both, galactic disks and their host DM halos provide multiple tracers of internally and externally-driven galaxy evolution. The second release of data from {\it Gaia} \citep{gaia18} cross-matched with SDSS (Sloan Digital Sky Survey) reveals radius-velocity phase-space correlations originated during mergers which occured over the last few Gyrs. Numerical simulations of Milky Way-type DM halos have predicted that such `streamers' will be present in the DM halos, and accumulate since $z\sim 1$ \citep{rom09}. Recent mixture model analysis of SDSS-{\it Gaia} DR2 catalog has confirmed this effect \citep{neci18}. Additional `archaeological' work will uncover other relics imprinted on the halo kinematics during its buildup history.

\section*{Acknowledgements}
We thank Phil Hopkins for providing us with the current version of GIZMO, and are grateful to Alar Toomre and Scott Tremaine for illuminating discussions. This work has been partially supported by the HST/STScI Theory grant AR-14584, and by JSPS KAKENHI grant \#16H02163 (to I.S.). I.S. is grateful for support from International Joint Research Promotion Program at Osaka University. The STScI is operated by the AURA, Inc., under NASA contract NAS5-26555. Simulations have been performed on the University of Kentucky DLX Cluster.


\end{document}